# Thickness Dependent Itinerant Ferromagnetism in Ultrathin Ba-Doped SrRuO$_3$ films

Zeeshan Ali[1*], Mohammad Saghayezhian[1], Prahald Siwakoti[1,2], Roshan Nepal[1], Jiandi Zhang[1,3,†]

[1]Department of Physics & Astronomy, Louisiana State University, Baton Rouge, LA 70803, USA.
[2]Department of Physics & Astronomy, University of Tennessee, Knoxville, TN 37996, USA.
[2]Beijing National Laboratory for Condensed Matter Physics, Institute of Physics, Chinese Academy of Sciences, Beijing 100190, People's Republic of China.

## Abstract

The electronic and magnetic properties in $AB$O$_3$ perovskite oxides are extremely sensitive to lattice structure but also the dimensionality, such as the thickness in thin film form. Here, we report the thickness-dependent electro-magnetic properties of ultrathin epitaxially stabilized Sr$_{1-x}$Ba$_x$RuO$_3$ ($x$ = 0.08, 0.2) thin films on SrTiO$_3$ (001) substrate. The results reveal that the Ba-doping ($0.08 \leq x \leq 0.20$) reduces RuO$_6$ orthorhombic distortions existing in SrRuO$_3$ and induces a tetragonal distortion as evidenced by out-of-plane lattice expansion. A metal-to-insulator transition, accompanied by a ferromagnetic to non-magnetic transition occurs with reducing film thickness from 10 to 3 unit-cell (u.c.) for both $x$ = 0.08 and 0.2, regardless of the doping level. The results suggest that the effects of compositional vacancies and surface/interface contributions introduced via dimensional confinement are more dominant than A-site chemical disorder or structural distortion for the loss of metallicity and ferromagnetism in ultra-thin epitaxial films.

**Key Words:** Ruthenates, Metal-Insulator Transition, Itinerant Ferromagnetism, Structural Distortion

[*]zee89ali@gmail.com
[†]jiandiz@iphy.ac.cn

## I. INTRODUCTION

Among complex oxides ($ABO_3$), it is well known that $B$-site transition-metal dictates the functional properties [1–3]. Minor modifications brought via changing octahedral environment, and bond symmetry can modify the intricate coupling between structure, and electronic spin degrees of freedom, leading to emergent properties [2,4]. One typical example is 4$d$ perovskite ruthenates ($A$RuO$_3$), where varying the $A$-site isovalent substitution from Ca to Sr to Ba not only triggers modified crystal structures but also affects magnetic ground states [5–10]. In Sr$_{1-x}$Ca$_x$RuO$_3$ bulk and thin film family, the compound retains orthorhombicity, though ferromagnetism ($x = 0$) evolves to paramagnetism ($x \approx 0.7$) due to altered Fermiology [5–8]. On the other hand, the Ba-placement at Sr-site drives a cubic structure but lowers the $T_C$ to 60 K for the final member: BaRuO$_3$ in bulk Sr$_{1-x}$Ba$_x$RuO$_3$ series [5,6]. For the Sr$_{1-x}$Ba$_x$RuO$_3$ thin-films, using strain engineering, the tuning of crystal symmetry from orthorhombic to cubic-like phase without octahedral rotations as well as modification of perpendicular magnetic anisotropy and magnetization dynamics is recently reported [11]. Such tuning of transition metal perovskite oxides for emergent functionalities via film epitaxy and heterostructure engineering is at the forefront of material science. Through epitaxial strain [12–20], interfacial effects [21–25], superlattices [26–30], confinement [31–35], doping/implantation [36–38], and dimensionality [39–43], a delicate equilibrium among structure, spin, charge, and orbital degrees of freedom could be manipulated, leading to novel properties that are different from the bulk.

Gaining insight into the nature of electronic and magnetic properties of the Sr$_{1-x}$Ba$_x$RuO$_3$ ($x$ = 0.08, 0.2) thin-films family in an ultra-thin regime, especially in a range of only a few unit-cell thicknesses, is the focus of this work. The Ba-doping into ultra-thin epitaxial SrRuO$_3$ lattice introduces a local RuO$_6$ octahedra elongation, while the nature of electron transport and itinerant magnetism shows stability against Ba-induced local-structural distortion. The placement of $Ba^{2+}$ ion at $Sr^{2+}$ site introduces randomized cation disorder; however, a thickness-driven metal-to-insulator transition (MIT) and non-magnetic state occurs at the same thickness of 3 unit-cells (u.c.) for both doping levels ($x$ = 0.08, 0.2). The results indicate that disordering effects such as compositional vacancies and surface/interface contributions introduced via dimensional confinement are more significant than the perturbative effects of local-structural distortion and cation disorder, leading to MIT and non-magnetic state.

## II. EXPERIMENTAL DETAILS

The $Sr_{1-x}Ba_xRuO_3$ ($x$ = 0.08, 0.2) thin films were grown by pulse laser deposition (PLD) using a KrF excimer laser with a 10 Hz repetition rate, and energy density ~1 J/cm². The $TiO_2$-terminated $SrTiO_3$ (001) substrates were employed, whereas thin films were grown at 700° C under an oxygen pressure of 100 mTorr and cooled down at 100 mTorr. The film thickness was monitored by an *in-situ* reflection high-energy electron diffraction (RHEED). The films crystal structure was characterized by using a PanAlytical X'Pert thin-film diffractometer. Electron transport measurements were performed via the Quantum Design Physical Property measurement system in a four-probe configuration. The sample magnetization was studied by using a Quantum Design Superconducting Quantum Interference Device, reciprocating sample option.

## III. RESULTS AND DISCUSSIONS

a) **Film Epitaxy**

**FIG. 1(a)** shows a schematic of the $Sr_{1-x}Ba_xRuO_3$ thin film epitaxy on a $TiO_2$-terminated $SrTiO_3$ (001) substrate, whereas **FIG. 1(b-c)** displays i*n-situ* reflective high energy electron diffraction (RHEED) data of 10 u.c. $Sr_{1-x}Ba_xRuO_3$ ($x$ = 0.08 and 0.2) films. **FIG. 1(b-c)** displays the RHEED images of both the substrate and films along with the RHEED spot intensities vs. growth time. The RHEED image of SrTiO$_3$ substrate [see inset of **FIG. 1(b-c)**] before deposition shows high-intensity specular spots and Kikuchi lines, verifying high substrate quality. **FIG. 1(b-c)** advocates that diffraction intensity of (0,0) spot oscillated in a layer-by-layer mode and as the film grows thicker, the oscillation intensity saturates, reflecting a preference toward step-flow epitaxy, leading to smooth thin films as established by after-growth specular diffraction spots [inset of **FIG. 1(b-c)**]. Overall, both the time-dependent intensity profile of RHEED spots and streaky RHEED pattern after film growth point to a flat crystalline surface having two-dimensional growth typically observed for ruthenates perovskites.

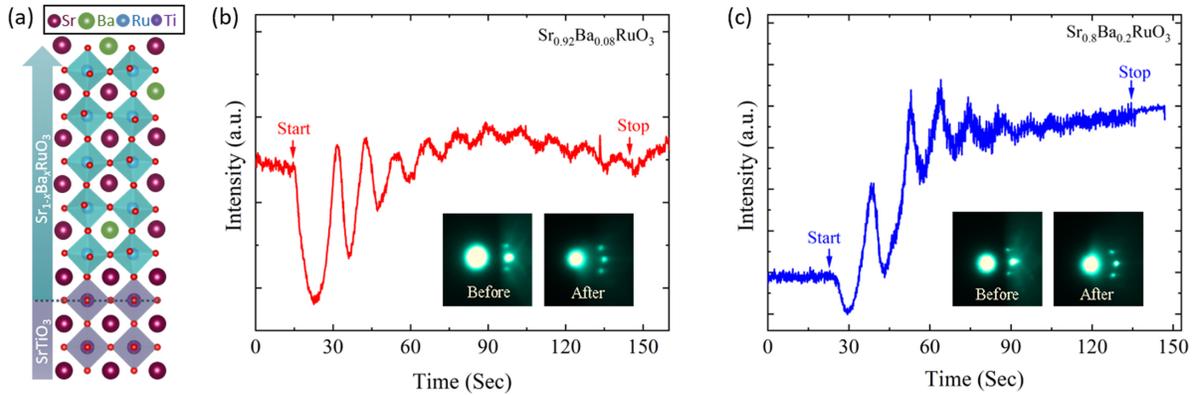

**FIG. 1.** (a) Schematic representation of Sr$_{1-x}$Ba$_x$RuO$_3$ film grown on SrTiO$_3$ (001) substrate. (b-c) Time-dependent high energy electron diffraction (RHEED) intensity profile of (0,0) spot for Sr$_{1-x}$Ba$_x$RuO$_3$ ($x$ = 0.08, 0.2) thin films with thickness of 10 unit-cells (u.c.). The inset in panel (b-c) shows the specular spots diffraction pattern before and after film growth.

b) **Crystal Structure**

The crystal structure of Sr$_{1-x}$Ba$_x$RuO$_3$ thin films was investigated via X-ray diffraction (XRD). **FIG. 2(a)** shows coupled (θ-2θ) X-ray scan around (002)-SrTiO$_3$ reflection for 10 u.c. Sr$_{1-x}$Ba$_x$RuO$_3$ ($x$ = 0.08, 0.2) thin films. The XRD spectra showed only substrate and film peaks confirming coherent film epitaxy and excellent crystalline quality [**FIG. 2(a)**]. The films maintain pseudocubic structural symmetry while the out-of-plane pseudocubic lattice parameter increases systematically with increasing Ba substitution from 0.08 to 0.2, i.e., a Ba-doping dependent tetragonal elongation compared to the SrRuO$_3$ films [11,19,20,36,44–46]. The Sr$_{0.92}$Ba$_{0.08}$RuO$_3$ film shows an out-of-plane lattice parameter of $c_{pc}$ = 4.029 ± 0.05 (bulk $c_{pc}$ = 3.944 Å), which enhances slightly to $c_{pc}$ = 4.076 ± 0.08 (bulk $c_{pc}$ = 3.968 Å) for Sr$_{0.8}$Ba$_{0.2}$RuO$_3$ film. The enhancement in $c_{pc}$ as compared to bulk counterparts is due to compressive strain, which enforces the in-plane parameter to undergo compression to match the SrTiO$_3$ (3.905 Å), while the out-of-plane constant increases.

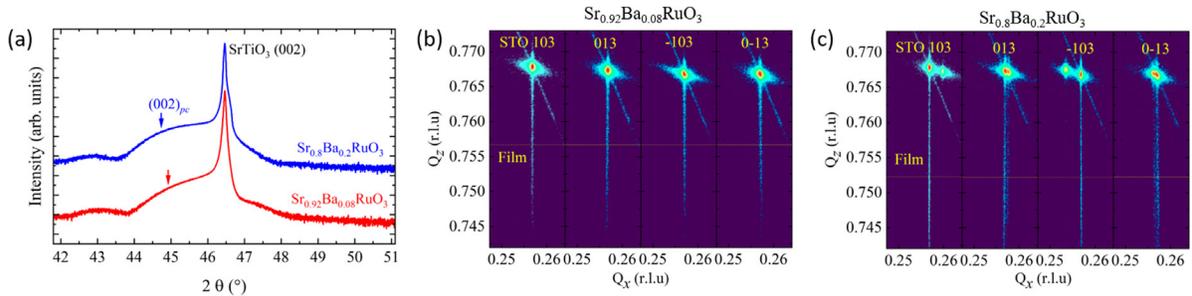

**FIG. 2.** (a) Coupled (θ-2θ) X-ray scan of Sr$_{1-x}$Ba$_x$RuO$_3$ ($x$ = 0.08, and 0.2) films around (002)-SrTiO$_3$. (b-c) Reciprocal space mapping about the (103)$_{pc}$ reflections for Sr$_{0.92}$Ba$_{0.08}$RuO$_3$ and Sr$_{0.8}$Ba$_{0.2}$RuO$_3$ films, respectively.

To further understand crystal structure, we have performed reciprocal space mapping (RSM) around (103)$_{pc}$ reflections for 10 u.c. Sr$_{0.92}$Ba$_{0.08}$RuO$_3$ and Sr$_{0.8}$Ba$_{0.2}$RuO$_3$ films [**FIG. 2(b-c)**]. The RSM is obtained around (103)$_{pc}$ SrTiO$_3$ reflection with a φ angle of 0°, 90°, 180°, and 270°. We observe that Sr$_{0.92}$Ba$_{0.08}$RuO$_3$ [**FIG. 2(b)**] film shows matching $Q_z$ reflection for family of (103)$_{pc}$ reflections, indicating a tetragonal symmetry [16,45,47]. Likewise, Sr$_{0.8}$Ba$_{0.2}$RuO$_3$ film also holds a tetragonal symmetry [**FIG. 2(c)**]. Moreover, the trend of identical $Q_x$ values as these of SrTiO$_3$ endorses the epitaxially strained nature of films, ensuring excellent quality.

**c)     Dimensionality-driven electric transport**

Resistivity as a function of temperature for thicknesses ranging from 10 to 3 u.c. for Sr$_{1-x}$Ba$_x$RuO$_3$ ($x$ = 0.08, and 0.2) series is shown in **FIG. 3(a-b)**. First, we observe that increasing the Ba-doping from 0.08 to 0.2 results in an increase of the resistivity value at room temperature for films with the same thickness (10, 7 u.c.), suggesting a cation-disorder effect. However, such resistivity enhancement posts an insignificant impact on $\rho(T)$ functional dependencies (more discussion later). More importantly, film thickness reduction results in a metal-to-insulator transition (MIT) at the same thickness of 3 u.c. for both doping levels ($x$ = 0.08, 0.2). We can divide the electron transport [**FIG. 3(a-b)**] into three electronic regimes as the film thickness is systematically varied for the Sr$_{1-x}$Ba$_x$RuO$_3$ ($x$ = 0.08, and 0.2) series. First, as shown in **FIG. 3(a)**, the 10 u.c. film of Sr$_{0.92}$Ba$_{0.08}$RuO$_3$, which has the lowest- room-temperature resistivity among the different thicknesses of the films, displays metallic behavior characterized by $d\rho/dT > 0$ down to the lowest temperature of 5 K. A clear kink in $\rho(T)$ curve appears at around ~ 133 K, concurring with the magnetic phase transition (see discussion later on), similar to that observed in the bulk crystal of SrRuO$_3$ [48,49]. The decreasing film thickness to an intermediate region (7 and 5 u.c.) triggers amplification in the room-temperature resistivity, whereas the films exhibit a metallic behavior only at high temperatures [**FIG. 3(a)**]. Specifically, the low-temperature resistivity exhibits negative slop ($d\rho/dT < 0$) below the resistivity minimum at ~ 23 K in 7 u.c. film, thus showing non-metallic character. The reduction of film thickness to 5 u.c. not only increases film resistivity but also shifts the resistivity minimum to the higher temperature of ~ 68 K. The existence of such resistivity upturn with a minimum is a sign that the system is at the edge of the metal-insulator transition [39–43,50]. Finally, when the film thickness is further reduced to 3 u.c., an insulating behavior ($d\rho/dT < 0$) is observed in the entire temperature range.

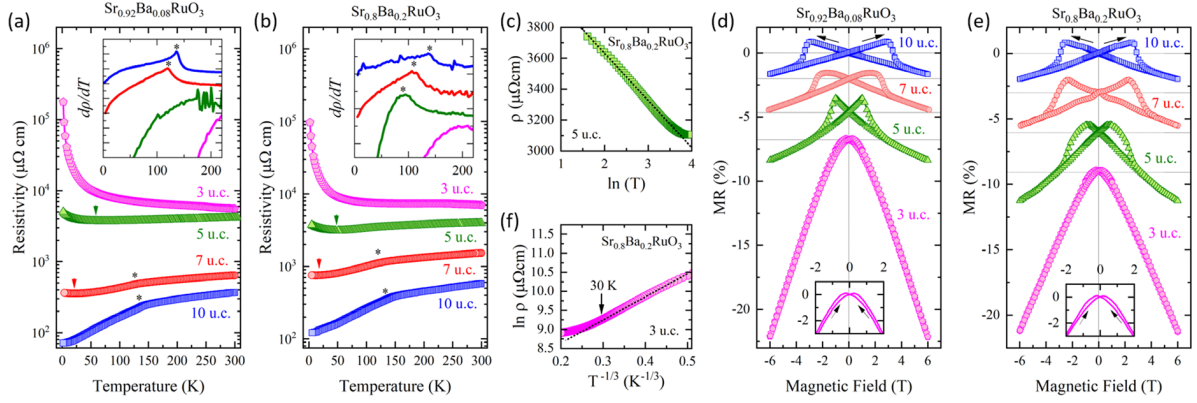

**FIG. 3.** (a-b) $Sr_{1-x}Ba_xRuO_3$ ($x = 0.08$, and 0.2) films resistivity as function of temperature for different thickness. The insets in panel (a-b) are temperature-dependent resistivity derivative ($d\rho/dT$) plotted against temperature ($T$). In panel (a-b), resistivity minimum temperature and $T_C$ are marked arrow and *, respectively. (c) Resistivity as function of temperature logarithm [$\ln(T)$] for 5 u.c. $Sr_{0.8}Ba_{0.2}RuO_3$ film. The black dotted-line is the linear fitting. Magnetoresistance (MR) of (d) $Sr_{0.92}Ba_{0.08}RuO_3$, and (e) $Sr_{0.8}Ba_{0.2}RuO_3$ films measured at 5 K with magnetic field applied along out-of-plane direction. The MR curves are off-set to avoid the overlap. The inset of (d-e) shows low-field MR of 3 u.c. $Sr_{1-x}Ba_{1-x}RuO_3$ ($x = 0.08$, and 0.2) highlighting the weak MR hysteresis. (f) Logarithm of resistivity plotted versus $T^{-1/3}$ (2D-Mott VRH) for 3 u.c. $Sr_{0.8}Ba_{0.2}RuO_3$ film. The black dotted-lines in (f) is the linear fitting.

The three kinds of transport regimes are also preserved in $Sr_{0.8}Ba_{0.2}RuO_3$ films, even though the increase in Ba-doping ($x = 0.2$) results in a minor increase in room-temperature resistivity [**FIG. 3(b)**]. Noticeably, a thickness-dependent MIT occurs at an identical thickness of 3 u.c. regardless of Ba-doping levels ($x = 0.08$ and 0.2). The 10 u.c. of $Sr_{0.8}Ba_{0.2}RuO_3$ film shows metallicity down to the lowest temperature, while the intermediate region (7 and 5 u.c.) films retain a high-temperature metallicity, yet resistivity upturn arises at low-temperature near ~ 19 K and ~ 55 K, respectively. Below this thickness, a fully insulating regime ascends as realized for 3 u.c. film. Thus, we conclude that the $Sr_{1-x}Ba_{1-x}RuO_3$ ($x = 0.08$, 0.2) series undergo a metal-insulator-transition (MIT) with films thinner than or equal to 3 u.c., regardless of Ba-doping level ($x = 0.08$, 0.2). This is analogous to the results of SrRuO$_3$ films grown on SrTiO$_3$ (001), where MIT occurs near 3 unit-cells [39–41].

The resistivity upturn is an indicator for MIT and has been observed in several oxide films [19,33,39–43,50–52]. In the ultrathin-film regime, the effect of weak localization induced via disorder could trigger resistivity upturn. The presence of disorder/defects results in the amplification of backscattering of the conducting electrons, leading to the interference of electronic wavefunctions and thus causing larger resistivity films [40–43,50–55]. The phenomenon of weak localization in two-dimensional systems results in a quantum correction to conductivity with a characteristic logarithmic [ln($T$)] dependence. Apparently, the signatures of weak localization-induced resistivity minima are prevailing in Sr$_{1-x}$Ba$_x$RuO$_3$ ($x$ = 0.08, and 0.2) series at a film thickness of 5 unit-cells. Specifically, the 5 u.c. Sr$_{0.8}$Ba$_{0.2}$RuO$_3$ film resistivity reveals a ln($T$) dependence in a low-temperature regime (< 50 K) as shown in **FIG. 3(c)**. Meanwhile, the effect of electron-electron correlations can also produce an ln($T$) dependence, considering the fact that electron-electron correlations might be enhanced in low-dimensional systems [10,33,56]. However, one can differentiate the effect of electron-electron correlations from weak localization via magnetoresistance measurements as we will discuss later.

The magnetoresistance (MR) of Sr$_{1-x}$Ba$_x$RuO$_3$ ($x$ = 0.08, 0.2) films is shown in **FIG. 3(d-e)**. The MR is defined as MR = {$\rho(H) - \rho(0)$}/$\rho(0)$, where $\rho(0)$ and $\rho(H)$ and are resistivities in the absence and presence of an external magnetic field, respectively. Here, the magnetic field $(H)$ is applied along the film-normal direction. For films under the study, both doping levels [Sr$_{1-x}$Ba$_{1-x}$RuO$_3$ ($x$ = 0.08, 0.2)] characteristically show identical MR behavior at 5 K. The MR of 10-5 u.c. ($x$ = 0.08, 0.2) films displays a "butter-fly" hysteretic-MR, a signature of FM ordering. In contrast, for the 3 u.c., the butter-fly MR becomes very weak (nearly vanishing), indicating a very weak-magnetic system [inset of **FIG. 3(d-e)**]. At the same time, the MR enhances monotonically with a reduction in film dimensionality, which could be associated with charge localization. Similar behavior has been observed in the CaRuO$_3$ films though without long-range magnetic ordering [35].

The magnetoresistance can be valuable to distinguish the contribution of weak localization from electron-electron correlation in resistivity upturn. The presence of the external magnetic-field presence can demolish the electronic wave-functions coherence by dephasing the two wave-functions that are traveling in clockwise and anticlockwise trajectories, leading to destructive interference, and abolishment of weak localization, hence a negative magnetoresistance [42,53,57–59]. In contrast, the electronic-correlation effect triggers a positive magnetoresistance [53,57,58]. As seen in **FIG. 3(d-e)**, the 5 u.c. films ($x$ = 0.08, 0.2) holding a resistivity minimum shows mainly negative magnetoresistance, confirming that resistivity minimum arises due to disorder-induced localization.

As mentioned before, the dimensional confinement in the Sr$_{1-x}$Ba$_x$RuO$_3$ ($x$ = 0.08, and 0.2) films results in a progressive evolution in the transport behavior from a metallic region (10-7 u.c.) to a weak localization state (5 u.c.) to a highly insulating behavior at 3 unit-cells. As shown in **FIG. 3(f)**, the low-temperature resistivity of insulating 3 u.c. Sr$_{0.8}$Ba$_{0.2}$RuO$_3$ film (taken as an example) exhibits two-dimensional variable range hopping (VRH) conduction characteristics: $\ln(\rho) \propto T^{-1/3}$. In the VRH conduction mechanism at low-temperature, the electrons can hop between localized states, leading to $\rho = \rho_0 \, exp\left[\left(\frac{T_0}{T}\right)^{\frac{1}{d+1}}\right]$, where $T_0$ is characteristic temperature signifying the degree of the disorder, and $d$ is system dimensionality ($d$ =1, 2, and 3 for one-, two-, and three-dimension, respectively) [60,61]. The results described above suggest that the observed thickness-dependent MIT is administered by disorder-driven Anderson localization [60–63]. Disorder refers to random fluctuations or variations due to the presence of impurities or defects in a system that breaks the electronic wave symmetry and causes the electronic wavefunctions to become localized, leading to the suppression of electrical conductivity. In our case here, the Ba-substitution introduces A-site cation disorder as well as local-structural distortion. At the same time, reducing film thickness can introduce and amplify additional disorder effects such as compositional (Ru and O) vacancies, and surface/interface contributions [34,38,46,64–67]. Compared to the chemical disorder at the A-site by Ba-doping, the disorder induced by condensing film thickness can be more severe in the activation of electron localization since, for example, the appearance of compositional (Ru and O) vacancies can drastically affect the *p-d* orbital hybridization in the building block $RuO_6$ of perovskite oxides. This is likely the reason that the films exhibit the same critical thickness for MIT regardless of different Ba doping levels [10,34,68,69].

### d) Thickness dependent magnetization

To study the magnetic properties of $Sr_{1-x}Ba_xRuO_3$ ($x$ = 0.08, 0.2) thin films, we have performed SQUID magnetometry for films with different thicknesses ($t$ = 3, 5, 7, 10 u.c.). For measuring the magnetization versus temperature [M($T$)], the films were cooled down under a field of 2000 Oe to 5 K, and then their M($T$) was measured during warm-up with a field of 100 Oe. **FIG. 4(a)** shows the M($T$) of $Sr_{0.92}Ba_{0.08}RuO_3$ films with thickness varying from 3 to 10 u.c. The M($T$) shows that thicker film (10 u.c.) holds a paramagnetic (PM) to ferromagnetism (FM) transition at 136 ± 4 K [almost the same as the kink position ~ 133 K in *ρ(T)* curve shown in FIG. 3(a)]. However, systematic reduction of film thickness suppresses the FM order and $T_C$, ending with the 3 u.c. film non-magnetic. These observations suggest that thickness-dependent transition from FM to non-FM state occurs at 3 u.c. thickness. Likewise, the M($T$) data of $Sr_{0.8}Ba_{0.2}RuO_3$ films with varying thickness exhibit similar behavior as that of $Sr_{0.92}Ba_{0.08}RuO_3$ films. Overall, the observations suggest that thickness-dependent transition from FM to non-FM state occurs at 3 u.c. thickness in both $Sr_{1-x}Ba_xRuO_3$ ($x$ = 0.08, 0.2) series.

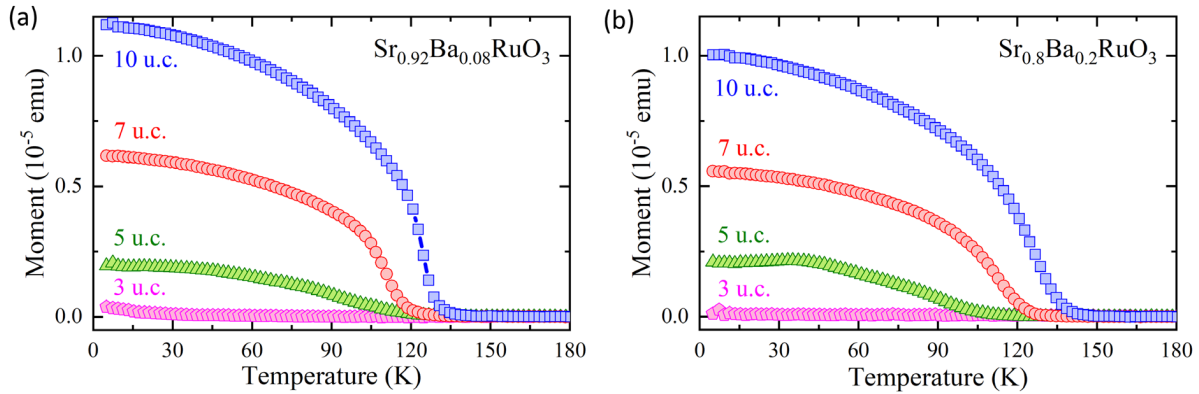

**FIG. 4.** Magnetization versus temperature [M($T$)] of (a) $Sr_{0.92}Ba_{0.08}RuO_3$ and (b) $Sr_{0.8}Ba_{0.2}RuO_3$ films measured along out-of-plane direction.

Lastly, we have plotted the Curie-temperature ($T_C$) and metal-insulator-transition temperature ($T_{MIT}$) as functions of film thickness for $Sr_{1-x}Ba_xRuO_3$ ($x$ = 0.08, 0.2) films [**FIG. 5**]. It could be seen that as the film thickness is reduced to 3 u.c., the FM order is diminished without a clear onset of magnetism (no measurable $T_C$ either) for both doping levels. At the same time, the transport data has shown that 3 u.c. films are insulating as revealed by metal-insulator-transition temperature ($T_{MIT}$) [**FIG. 5**]. Thus, the emergence of a non-FM state coincides with the insulating state. The $T_C$ in a thicker film regime is found to be nearly constant. This implies that the metallicity and FM ordering go parallel to each other, which is expected for an itinerant system such as $Sr_{1-x}Ba_xRuO_3$ ($x$ = 0.08, 0.2). The trend of $T_C$ is also consistent with the reports on $SrRuO_3$ films, where films are found to be non-magnetic and insulating near 2-3 unit-cells [39–41,70].

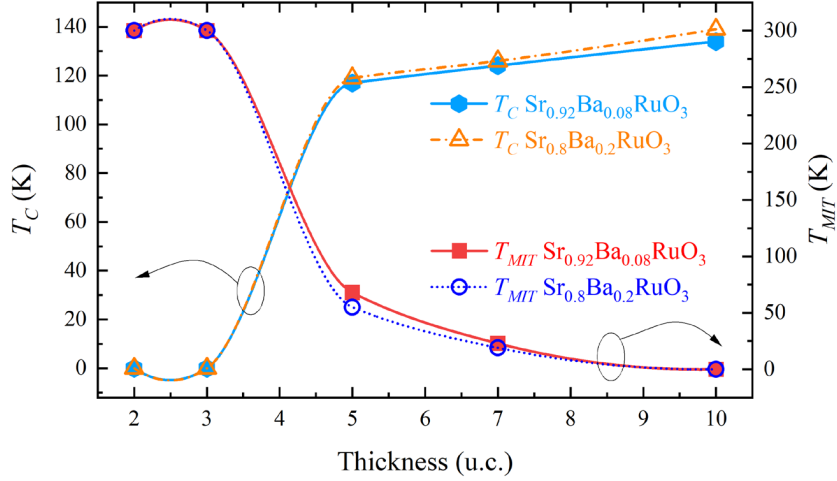

**FIG. 5.** Curie-temperature ($T_C$), and metal-insulator transition temperature ($T_{MIT}$) plotted against film thickness for series of $Sr_{0.92}Ba_{0.08}RuO_3$, and $Sr_{0.8}Ba_{0.2}RuO_3$ films.

## IV.  CONCLUSION

To summarize, we have studied the thickness-dependent electro-magnetic properties of ultrathin $Sr_{1-x}Ba_xRuO_3$ ($x$ = 0.08, 0.2) thin films grown on SrTiO$_3$ (001). The study reveals that Ba-doping induces out-of-plane lattice expansion, yet itinerant magnetism of SrRuO$_3$ displays robustness against Ba-doping. Notably, the electron transport reveals that Ba-doped SrRuO$_3$ films ($x$ = 0.08, 0.2) undergo an MIT at 3-unit cells, while above this thickness, films are metallic. The emergence of an insulating state coincides with a non-FM state. The outcomes highlight that reducing film thickness introduces and amplifies the disordering effects such as compositional (Ru and O) variation, and surface/interface contributions, which are dominant compared to the Ba-induced effects of cation-disorder and local-structural distortion, triggering the metal-insulator and ferromagnetic-nonmagnetic transitions in ultrathin epitaxial films.

# ACKNOWLEDGMENTS

This work is primarily supported by the US Department of Energy (DOE) under Grant No. DOE DE-SC0002136.